\documentclass[doublecol]{epl2}
\usepackage{tipa}
\usepackage{pifont}
\usepackage{txfonts}
\usepackage{amssymb}
\usepackage{graphicx}
\usepackage{amsfonts}
\usepackage{epsfig}

\institute{ Department of Physics and ITP, The Chinese University of Hong Kong,
Hong Kong, China}

\abstract{We analyze ground-state behaviors of fidelity susceptibility (FS) and
show that the FS has its own distinct dimension instead of real system's
dimension in general quantum phases. The scaling relation of the FS in quantum
phase transitions (QPTs) is then established on more general grounds. Depending
on whether the FS's dimensions of two neighboring quantum phases are the same
or not, we are able to classify QPTs into two distinct types. For the latter
type, the change in the FS's dimension is a characteristic that separates two
phases. As a non-trivial application to the Kitaev honeycomb model, we find
that the FS is proportional to $L^2\ln L$ in the gapless phase, while $L^2$ in
the gapped phase. Therefore, the extra dimension of $\ln L$ can be used as a
characteristic of the gapless phase.}

\begin{document}

\title{Scaling dimension of fidelity susceptibility in quantum phase transitions}
\author{Shi-Jian Gu\footnote{Correspondence: sjgu@phy.cuhk.edu.hk} \and Hai-Qing Lin
}

\maketitle

\shortauthor{S. J. Gu \etal}




{\it Introduction ---} Let us consider a general QPT\cite{Sachdev} occurring in
the ground state $|\Psi _{0}(\lambda )\rangle $ of a quantum many-body
Hamiltonian
\begin{equation}
H=H_{0}+\lambda H_{I},  \label{eq:Hamiltonian}
\end{equation}%
where $H_{I}$ is the driving Hamiltonian and $\lambda $ denotes its
strength. The motivation of the fidelity approach \cite{Nielsen1} to QPTs is
that the overlap between two ground states separated by a small amount $%
\delta \lambda $, i.e. $F(\lambda ,\lambda +\delta \lambda )=|\langle \Psi
_{0}(\lambda )|\Psi _{0}(\lambda +\delta \lambda )\rangle |$, is expected to
show a minimum around the critical point $\lambda _{c}$ due to the dramatic
change in structures of the ground-state wavefunction \cite%
{HTQuan2006,Zanardi06}. This interesting insight to QPTs from quantum
information theory \cite{Nielsen1} had then been demonstrated in a few
strongly correlated systems \cite%
{PZanardi0606130,MCozzini07,MCozzini072,Buonsante1}. It was realized
consequently that the leading term of the fidelity, called the FS\cite%
{WLYou07} or the Riemannian metric tensor\cite{PZanardi0701061}, should play
a key role in such a new approach to QPTs. After that, various issues based
on the fidelity or its leading term \cite%
{HQZhou0701,HQZhou07042940,LCVenuti07,SJGu072,SChen07,WXG,MFYang07,HMKwok07,NPaunkovic07,XWang08032940,HMKwok08,JMa08,AHamma07,JHZhao0803,SYang08,DFAbasto08,HTQuan}
, including scaling and universality class \cite{LCVenuti07,SJGu072}, and
its role in topological QPTs \cite{AHamma07,JHZhao0803,SYang08,DFAbasto08}
etc, were raised and addressed.

However, it seems to us that all relevant studies took it for granted that
the FS, in general quantum phases, has dimension $d$, i.e.,$\chi _{F}\propto
L^{d}$ where $d(L)$ is the dimension (length) of the system. In this work,
we will show instead that the FS has its own dimension depending on both the
scaling dimension of the driving Hamiltonian and long-range behaviors of
their correlations. The distinct dimension of the FS means that, in a class
of quantum phases, the adiabatic response of the ground state to driving
parameter is no longer proportional to the system size. This property is
different from our previous understanding of critical phenomena from
statistical quantities, such as energy that is extensive in a thermodynamic
system. Therefore, the observation not only provides a new angle to
understand the role of FS in QPTs, but also is of fundamental importance to
apply quantum adiabatic theorem to scalable systems.

The critical exponents of the FS and their scaling relation then will be
proposed in a more general way. Clearly, the FS's dimension can be changed, or
not changed in QPTs, this property classifies all QPTs into two different
types. For a class of QPTs, the change in the FS's dimension is a
characteristic that separates two phases. As a non-trivial application, we will
show that, in the gapless phase of the Kitaev honeycomb model, the dimension of
the FS becomes $L^{2}\ln L$ rather than $L^{2}$ in the gapped phase. Therefore,
the extra dimension of $\ln L$ becomes a characteristic of the gapless phase.

{\it Scaling analysis revisited ---} The FS is defined as the leading term of
the fidelity $\chi _{F}=-\lim_{\delta \lambda\rightarrow 0}2\ln F/(\delta
\lambda )^{2}$. For the Hamiltonian system [Eq. (\ref{eq:Hamiltonian})], the
ground-state FS can be evaluated from \cite{WLYou07,PZanardi0701061}
\begin{equation}
\chi _{F}(\lambda )=\sum_{n\neq 0}\frac{|\langle \Psi _{n}(\lambda
)|H_{I}|\Psi _{0}(\lambda )\rangle |^{2}}{[E_{n}(\lambda )-E_{0}(\lambda
)]^{2}},  \label{eq:FSperturbation}
\end{equation}%
where $H(\lambda )|\Psi _{n}(\lambda )\rangle =E_{n}|\Psi _{n}(\lambda
)\rangle $. Here we would like to point out that, in the low-energy spectra
of the Hamiltonian, only those excitations with nonzero $\langle \Psi
_{n}(\lambda )|H_{I}|\Psi _{0}(\lambda )\rangle $ have contribution to the
FS. So if we define $\Delta $ as the energy gap between the ground state and
the lowest excitation with a nonzero $\langle \Psi _{n}(\lambda )|H_{I}|\Psi
_{0}(\lambda )\rangle $, the FS satisfies the following inequalities
\begin{eqnarray}
\chi _{F} &\leq &\frac{1}{\Delta }\sum_{n\neq 0}\frac{|\langle \Psi
_{n}(\lambda )|H_{I}|\Psi _{0}(\lambda )\rangle |^{2}}{[E_{n}(\lambda
)-E_{0}(\lambda )]}=-\frac{1}{2\Delta }\frac{\partial ^{2}E(\lambda )}{%
\partial \lambda ^{2}}  \nonumber \\
&\leq &\Delta ^{-2}[\langle \Psi _{0}(\lambda )|H_{I}^{2}|\Psi _{0}(\lambda
)\rangle -\langle \Psi _{0}(\lambda )|H_{I}|\Psi _{0}(\lambda )\rangle ^{2}].
\nonumber  \label{eq:inequalities}
\end{eqnarray}
Let us suppose $H_{I}=\sum_{r}V(r)$ and $N=L^{d}$. 1) If the system is
gapped, the FS has the same (or smaller) dependence on the system size as
the second derivative of the ground-state energy. 2) If $\chi _{F}/N$ still
increases with the system size, the ground state must be gapless.

The above inequalities can only tell us some qualitative information. To
find the scaling relation, we resort to the definition of the FS in terms of
correlation functions \cite{WLYou07},
\begin{equation}
\frac{\chi _{F}}{L^{d}}=\sum_{r}\int \tau G(r,\tau )d\tau ,
\label{eq:FScorrelation}
\end{equation}%
where $G(r,\tau )=\langle V(r,\tau )V(0,0)\rangle -\langle V(r,0)\rangle
\langle V(0,0)\rangle $. Under the scaling transformation
\cite{MAContinentinob} $r^{\prime }=s\,r,\tau ^{\prime }=s^{\zeta }\tau
,V(r^{\prime })=s^{-\Delta _{V}}V(r)$, it has been found that the FS scales
like $\chi _{F}/L^{d}\propto L^{d+2\zeta -2\Delta _{V}}$ \cite{LCVenuti07}
where $\zeta $ is the dynamic exponent and $\Delta _{V}$ is the scaling
dimension of $V(r)$ at the critical point, and $\chi _{F}$ is believed to be
extensive away from the critical point. The latter \textquotedblleft common
sense" is also the reason that, in the most previous studied, the FS is
believed to have the same critical exponent ${\alpha }$ on both sides of the
critical point, that is $\chi _{F}/L^{d}\propto |\lambda -\lambda
_{c}|^{-\alpha }$. However, it is not universally true.

For a general $d$-dimensional system, the correlation function $G(r,\tau )$
under the scaling transformation, becomes
\begin{eqnarray}
G(r^{\prime },\tau ^{\prime }) &=&s^{2(\Delta _{\lambda }-d)}G(r,\tau ) \\
&=&G(s\,r,s^{\zeta }\tau )
\end{eqnarray}%
where $\Delta _{\lambda }$ comes from the transformation $\lambda ^{\prime
}=s^{-\Delta _{\lambda }}\,\lambda $. Let $s=\tau ^{-1/\zeta }$, we then
have
\begin{equation}
G(r,\tau )=\tau ^{2(\Delta _{\lambda }-d)/\zeta }G(\,r\tau ^{-1/\zeta },1).
\end{equation}%
If we rearrange the expression,
\begin{equation}
G(\,r\tau ^{-1/\zeta },1)=[r\tau ^{-1/\zeta }]^{2(\Delta _{\lambda
}-d)}f(r\tau ^{-1/\zeta }).
\end{equation}%
which defines the scaling function $f(r\tau ^{-1/\zeta })$. The correlation
function becomes%
\begin{equation}
G(r,\tau )=\frac{1}{r^{2\Delta _{V}}}f(r\tau ^{-1/\zeta })
\end{equation}%
where $\Delta _{V}=d-\Delta _{\lambda }$ is just the scaling dimension of $%
V(r)$. If the correlation length is divergent, the correlation function
decays algebraically, and
\begin{equation}
\frac{\chi _{F}}{L^{d}}\sim \left\{
\begin{array}{cc}
L^{d+2\zeta -2\Delta _{V}}, & 2\Delta _{V}-2\zeta \neq d \\
{\rm ln}L, & 2\Delta _{V}-2\zeta = d
\end{array}%
\right. .
\end{equation}
For some collective systems, such as the Lipkin-Meshkov-Glick (LMG) model
\cite{Lipkin1965,JVidal05}, we will show below that the scaling dimension of
$\sigma ^{z}$ in the polarized phase is $N^{-1}$, hence $\chi _{F}$ is
intensive. Therefore, the FS has its own distinct dimension instead of real
system's dimension. The dimension is related to the size-dependence of the
fidelity of a given driving Hamiltonian, and as we show in our another paper
that $L^{d_{a}}$ actually determines the duration time scale in the quantum
adiabatic theorem \cite{SJGuQAD}, we will call it \textit{quantum adiabatic
dimension}(QAD), and use $d_{a}$ to denote it hereafter.

Therefore, to judge a quantum criticality from the FS, the scaling relation
and conditions should be revised. If the dimension of $\chi _{a}$ is $%
d_{a}^{+(-)}$ above (below) the critical point, then the rescaled FS, i.e. $%
\chi _{F}/L^{d_{F}^{\pm }}$, as an intensive quantity, scales like
\begin{equation}
\frac{\chi _{F}}{L^{d_{a}^{\pm }}}\propto \frac{1}{|\lambda -\lambda
_{c}|^{\alpha ^{\pm }}},  \label{eq:criticalbehav}
\end{equation}%
in the thermodynamic limit. Since the FS usually shows a maximum around the
critical point and scales like $\chi _{F}|_{\lambda =\lambda _{c}}\propto
L^{\mu }$ $(\mu =2d+2\zeta -2\Delta _{V})$\cite{LCVenuti07}, the general
scaling relation should be
\begin{equation}
\alpha ^{\pm }=\frac{\mu -d_{a}^{\pm }}{\nu }.
\end{equation}%
Here $\nu$ is the critical exponent of the correlation length, which defines
the length scale of the system around the critical point. In case the exponent
$\nu $ is also different (though we did not find such a case yet), a more
general relation is $\alpha ^{\pm }={(\mu -d_{a}^{\pm })}/{\nu ^{\pm }}$. Hence
the condition of singularity in the FS around the critical point is
\begin{equation}
\mu \geq d_{a}^{\pm }  \label{eq:criticalcondition}
\end{equation}%
instead of $\mu \geq d$ \cite{LCVenuti07}. A further remark is that, even if
the equality condition of Eq. (\ref{eq:criticalcondition}) is satisfied, it
is still possible for the system to undergo a logarithmic divergence at one
side of the critical point, i.e.
\begin{equation}
\frac{\chi _{F}}{L^{d_{a}}}\propto \ln |\lambda -\lambda _{c}|,
\label{eq:logdivergend}
\end{equation}%
if $\chi _{F}/L^{d_{a}}|_{\lambda =\lambda _{c}}$ diverges as $\ln L$.
Therefore, if and only if the QAD (including logarithmic dependence on the
system size) of the FS are the same above, at, and below the critical point $%
\lambda _{c}$, it is firm to say that the FS does not have singular behavior
around the critical point.

Eqs (\ref{eq:criticalbehav}-\ref{eq:logdivergend}) represent our first main
result of this work. These key relations define also a criteria to judge
quantum phase transitions in perspective of the FS. They are a
generalization of the results by Venuti and Zanardi \cite{LCVenuti07}. In
short, in order to study the scaling behavior of the FS, one should consider
the QAD of the driving Hamiltonian in the corresponding quantum phases. The
QAD can be generalized to a Hamiltonian defined in a high-dimensional
parameter space. In this case, different driving Hamiltonian might have
different QAD. In the following, we will check the validity of about
analysis in a well studied QPT of Landau's type and a topological QPT.

{\it Examples: the LMG model and the Kitaev honeycomb model---}
We first check the above scaling relations in an exactly solvable model\cite%
{Lipkin1965,JVidal05}, i.e. LMG model. Its Hamiltonian reads
\begin{equation}
H_{{\rm LMG}}=-\frac{1}{N}\sum\limits_{i<j}{\left( {\sigma _{x}^{i}\sigma
_{x}^{j}+\gamma \sigma _{y}^{i}\sigma _{y}^{j}}\right) }-h\sum\limits_{i}{%
\sigma _{z}^{i}},
\end{equation}%
where ${\gamma }$ denotes the anisotropicity. The prefactor $1/N$ is to
ensure a finite energy per spin. In the thermodynamic limit, the ground
state of the system for ${\gamma \neq 1}$ undergoes a second-order QPT at $%
h_{c}=1$. If $h>h_{c}$, the system is fully polarized, while if $0<h<h_{c}$
it is a symmetry-broken state. The ground-state fidelity of the LMG model
has been addressed previously by Kwok \textit{et al }\cite{HMKwok07}. They
found that $\chi _{F}$ is proportional to the system size $\mathit{N}$ in
the symmetry-broken phase, while is intensive in the fully polarized phase.
This difference makes that critical exponents to be different at both sides
of the critical point. Kwok \textit{et al }\cite{HMKwok07} obtained that $%
d_{a}^{+}=0$ for $h>1$, $d_{a}^{-}=1$ for $h<1$, $\mu =4/3$ at $h_{c}=1$,
and the correlation length exponent $\nu =2/3$, then the critical exponent
above $h_{c}$ is $\alpha ^{+}=(\mu -d_{a}^{+})/\nu =2,$ while below $h_{c}$
is $\alpha ^{-}=(\mu -d_{a}^{-})/\nu =1/2$.

The explicit differences of $d_{a}^{\pm }$ and $\alpha ^{\pm }$ in two
quantum phases of the LMG model has already been a straightforward
demonstration that the FS has its own dimension. According to Eq. (\ref%
{eq:FScorrelation}), the deep reason of such a difference is that ${\sigma
_{z}^{j}}\sim N^{-1}$ in region $h>1$, while ${\sigma _{z}^{j}}\sim N^{-1/2}$
if $0<h<1$, which leads to that the FS is intensive in the polarized phase,
while extensive in the symmetry-broken phase.

\begin{figure}[tbp]
\includegraphics[width=8cm]{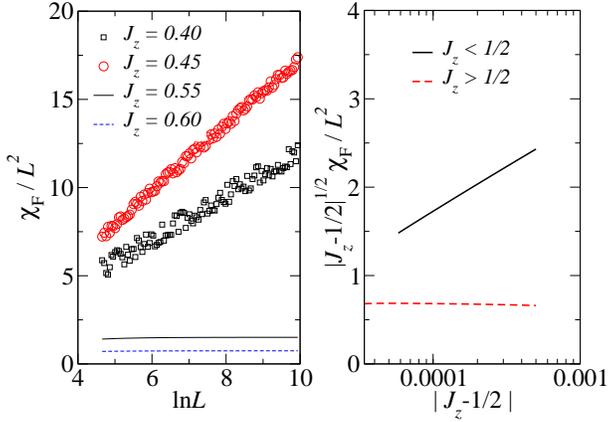}
\caption{(Color online) LEFT: The FS $\protect\chi_F/L^2$ as a function of $%
\ln L$ for the Kitaev model along the evolution line $J_x=J_y$ defined on $%
J_x+J_y+J_z=1$ plane. In cases of $J_z<1/2$, the explicit fluctuation of the
FS is due to appearances of infinite critical lines as the system size
increases \protect\cite{Kitaev,XYFeng07}. Nevertheless, the fluctuations do
not affect the logarithmic divergence of $\protect\chi_F/L^2$. RIGHT: The
rescaled $|J_z-1/2|^{1/2}\protect\chi_F/L^2$ as a function of $\ln|J_z-1/2|$
for a system of $L=20001$.}
\label{figure_kitaevdim.eps}
\end{figure}

As a non-trivial application, we study the FS in a topological QPT \cite%
{wen-book} occurring in the ground state of the Kitaev honeycomb model \cite%
{Kitaev}. The model is associated with a system of 1/2 spins which are
located at the vertices of a honeycomb lattice. The Hamiltonian reads:
\begin{equation}
H=-J_{x}\sum_{x{\rm bonds}}\sigma _{j}^{x}\sigma _{k}^{x}-J_{y}\sum_{y{\rm
bonds}}\sigma _{j}^{y}\sigma _{k}^{y}-J_{z}\sum_{z{\rm bonds}}\sigma
_{j}^{z}\sigma _{k}^{z},  \label{eq:HamiltonianKitaev}
\end{equation}%
where $j,k$ denote two ends of the corresponding bond linked to a vertex,
and $J_{\kappa }$ ($\kappa =x,y,z$) are coupling constants. The ground state
of the Kitaev honeycomb model consists of two different phases \cite{Kitaev}%
, i.e, a gapped phase with Abelian anyonic excitations and a gapless phase
with non-Abelian anyonic excitations. The critical behavior of the fidelity
in the model was previously addressed by two groups \cite{JHZhao0803,SYang08}%
. None of them addressed the QAD of the FS, but took it for granted that $%
d_{a}=d$. We define the phase diagram on the plane $J_{x}+J_{y}+J_{z}=1$,
and consider a certain line $J_{x}=J_{y}$ along which the ground state of
the system evolves at zero temperature. In this case, $J_{z}$ is the only
driving parameter. The Hamiltonian can be explicitly diagonalized in the
flux-free subspace \cite{Kitaev,XYFeng07} and the ground-state FS for a
system of $N=2L^{2}$(odd $L$) sites is \cite{SYang08}
\begin{equation}
\chi _{F}=\frac{1}{16}\sum_{\mathbf{q}}\left[ \frac{\sin q_{x}+\sin q_{y}}{%
\epsilon _{\mathbf{q}}^{2}+\Delta _{\mathbf{q}}^{2}}\right] ^{2}
\end{equation}%
where $q_{x\left( y\right) }=2n\pi /L,n=-(L-1)/2,\cdots ,(L-1)/2$, and
\begin{eqnarray}
\epsilon _{\mathbf{q}} &=&J_{x}\cos q_{x}+J_{y}\cos q_{y}+J_{z},  \nonumber \\
\Delta _{\mathbf{q}} &=&J_{x}\sin q_{x}+J_{y}\sin q_{y}.
\end{eqnarray}%
In the thermodynamic limit, the FS becomes
\begin{equation}
\frac{\chi _{F}}{L^{2}}=\frac{1}{64\pi ^{2}}\int_{-\pi }^{\pi
}dq_{x}\int_{-\pi }^{\pi }dq_{y}\left[ \frac{\sin q_{x}+\sin q_{y}}{\epsilon
_{\mathbf{q}}^{2}+\Delta _{\mathbf{q}}^{2}}\right] ^{2}.
\label{eq:fsintegration}
\end{equation}

\begin{table*}[tbp]
\caption{Critical exponents $\protect\mu ,\protect\nu ,d_{F}^{\pm },\protect%
\alpha ^{\pm }$, for the Ising model \protect\cite{Zanardi06}, the Kitaev
toric model(KTM) \protect\cite{DFAbasto08}, the LMG model \protect\cite%
{HMKwok07}, and the Kitaev honeycomb model (KHM).} \label{tab:critcalexp}
\begin{center}
\begin{tabular}{c|cc|cc|cc}
\hline
 Model & $\mu$ & $\nu$ & $d_a^+$ & $\alpha^+$ & $d_a^-$ & $\alpha^-$ \\
\hline 1D Ising model($h_c=1$) & 2 & 1 & 1 & 1
& 1 & 1 \\
\hline KTM($\lambda_{c} =\frac{1}{2}{\rm ln}(\sqrt{2}+1)$) & ln & 1 & 1 & ln
& 1 & ln \\
\hline LMG model($h_c=1$) & 4/3 & 2/3 & 0 & 2
& 1 & 1/2 \\
\hline KHM($J_c=1/2$) & 5/2 & 1 & 2 & 1/2 & 2+ln & 1/2-ln \\ \hline
\end{tabular}
\end{center}
\end{table*}

In the gapped phase $J_{z}>1/2$, $\epsilon _{\mathbf{q}}^{2}+\Delta _{%
\mathbf{q}}^{2}>0$. So there is no pole in the integrand of Eq. (\ref%
{eq:fsintegration}). $\chi _{F}/L^{2}$ is intensive and the QAD of the FS is
$d_{a}^{+}=2$. However, in the gapless phase, $\epsilon _{\mathbf{q}%
}^{2}+\Delta _{\mathbf{q}}^{2}$ has zero points in $k$ space. Expanding the
integration around the pole of the integrand, we find $\chi
_{F}/L^{2}\propto \int_{\pi /L}^{\Lambda }\frac{1}{k}dk\sim \ln L,$where $%
\Lambda $ is a cut-off. Therefore, $\chi _{F}/L^{2}$ manifests distinct size
dependence in the gapped and gapless phases. As a numerical demonstration,
we show the FS $\chi _{F}/L^{2}$ as a function of $\ln L$ in Fig. \ref%
{figure_kitaevdim.eps}(left) for various $J_{z}$. Obviously, $\chi
_{F}/L^{2} $ does not change as the system size increases in the gapped
phase, but is proportional to $\ln L$ in the gapless phase. At the critical
point $J_{z}=1/2$, it has been shown that $\chi _{F}\propto L^{5/2}$\cite%
{SYang08}. Therefore, according to the scaling analysis, the critical
exponents of the FS should be different at both sides of $J_{z}=1/2$. At $%
J_{z}=(1/2)^{+}$ it has been already obtained $\alpha ^{+}\simeq 0.5$, here
we find that%
\begin{equation}
\frac{\chi _{F}|J_{z}-J_{z,c}|^{1/2}}{L^{2}\ln L}\sim \ln |J-J_{z,c}|,
\end{equation}%
at $J_{z}=(1/2)^{-}$ [simplified as 1/2-ln for QAD and shown in Fig. \ref%
{figure_kitaevdim.eps}(right)]. Moreover, differ from both the quantum Ising
model and the LMG model, the FS has higher dimension 2+ln (for $\chi
_{F}\propto L^{2}\ln L)$ than the system's dimension 2 in the gapless phase.
The extra dimension of $\ln L$ has a special meaning because it only exists in
the gapless phase with non-Abelian anyonic excitations. Therefore, it can be
used as an characteristic of the gapless phase in the Kitaev honeycomb model.

From the point view of Eq. (\ref{eq:FScorrelation}), the deep reason behind
the extra dimension $\ln L$ is that the bond-bond correlation function of
the $z$-bond in Eq. (\ref{eq:HamiltonianKitaev}) decays exponentially in the
gapped phase, while algebraically, i.e., $G(r,0)\sim 1/r^{4}$, in the
gapless phase \cite{SYang08}. Therefore, the FS in Eq. (\ref%
{eq:FScorrelation}) becomes
\begin{equation}
\frac{\chi _{F}}{L^{d}}\sim \sum_{r}\frac{1}{r^{4-2\zeta }}.
\end{equation}%
The $\ln L$ divergence of the FS $\chi _{F}/L^{d}$ means that the dynamic
exponent should be $\zeta =1$.

Finally, we summarize the critical exponents of the FS in the above two typical
models, the quantum Ising model, and the Kitaev toric model(KTM) in Table. I.
The critical exponents of the FS in the latter two models are referred from Ref
\cite{Zanardi06} and \cite{DFAbasto08}, respectively. The table can be
prolonged as one consider more and more QPTs. No matter what transitions are
included, the QPTs can be classified into two distinct classes. For the first
class, the FS has the same QAD in the both quantum phases, such as the quantum
Ising model, while for the another class, the FS has different QAD, such the
LMG model and the Kitaev honeycomb model. Moreover, for the Kitaev honeycomb
model, we can see that the QAD actually plays a role of characteristic that
separates two quantum phases. These conclusions constitute our second main
result.

{\it A brief summary and a challenge---} In summary, we have analyzed the
dimension of the FS in quantum phase transitions. We have shown that the QAD of
the driving Hamiltonian is not always the same as the system's dimension. The
scaling relation of the FS in various QPTs is established on more general
grounds. The FS might have distinct critical exponent at both sides of the
critical point. So the QPTs can be divided into two classes based on the
criteria if the QAD is changed or not during the phase transition. Our results
also show that the QAD can be used as a characteristic that separates two
topological phases in the Kitaev honeycomb model. Therefore, the QAD provides a
quite distinct tool instead of the traditional order parameter to study QPTs.

Clearly, the QAD provides a unique classification of quantum phases. From this
point of view, a challenging problem might be that, for a $d$-dimensional
system, does there exist such a quantum phase with a QAD of $d+1$?

This work is supported by the Earmarked Grant for Research from the Research
Grants Council of HKSAR, China (Project No. CUHK 400807) and the Direct grant
of CUHK (A/C 2060344).


\begin{thebibliography}{99}
\bibitem{Sachdev}
  \Name{Sachdev S.}
  \Book{Quantum Phase Transitions}
  \Publ{Cambridge
University Press, Cambridge, UK}
  \Year{2000}.




\bibitem{Nielsen1}\Name{Nilesen M. A. and Chuang I. L.}
  \Book{Quantum Computation and Quantum Information}
  \Publ{Cambridge
University Press, Cambridge, UK}
  \Year{2000}.


\bibitem{HTQuan2006} \Name{Quan H. T., Song Z., Liu X. F., Zanardi P., and Sun C. P.}
\REVIEW{Phys. Rev. Lett.}{96}{2006}{140604}.


\bibitem{Zanardi06} \Name{Zanardi P. and Paunkovic N.} \REVIEW{Phys. Rev. E}{74}{2006}{031123}.

\bibitem{PZanardi0606130} \Name{Zanardi P., Cozzini M., and Giorda P.} \REVIEW{J. Stat.
Mech.}{2}{2007}{L02002}.

\bibitem{MCozzini07} \Name{Cozzini M., Giorda P., and Zanardi P.} \REVIEW{Phys. Rev. B}{75}{2007}{014439}.


\bibitem{MCozzini072}\Name{Cozzini M., Ionicioiu R., and Zanardi P.} \REVIEW{Phys. Rev.
B}{76}{2007}{104420}.

\bibitem{Buonsante1}\Name{Buonsante P. and Vezzani A.} \REVIEW{Phys. Rev. Lett.}{98}{2007}{110601}.


\bibitem{WLYou07}\Name{You W. L., Li Y. W., and Gu S. J.} \REVIEW{Phys. Rev. E}{76}{2007}{022101}.


\bibitem{PZanardi0701061}\Name{Zanardi P., Giorda P., and Cozzini M.} \REVIEW{Phys. Rev.
Lett.}{99}{2007}{100603}.




\bibitem{HQZhou0701}\Name{Zhou H. Q. and Barjaktarevic J. P.} \REVIEW{J. Phys. A: Math.
Theor.}{41}{2008}{412001}.

\bibitem{HQZhou07042940}\Name{Zhou H. Q., Zhao J. H., and Li B.} \REVIEW{J. Phys. A: Math. Theor.}
{41}{2008}{492002}.



\bibitem{LCVenuti07}\Name{Venuti L. C. and Zanardi P.} \REVIEW{Phys. Rev. Lett.}{99}{2007}{095701}.


\bibitem{SJGu072}\Name{Gu S. J., Kwok H. M., Ning, W. Q. and Lin H. Q.} \REVIEW{Phys.
Rev. B}{77}{2008}{245109}.

\bibitem{SChen07}\Name{Chen S., Wang L., Hao Y., and Wang Y.} \REVIEW{Phys. Rev. A}{77}{2008}{032111}.


\bibitem{WXG}\Name{Zanardi P., Quan H. T., Wang X. G., and Sun C. P.} \REVIEW{Phys. Rev.
A }{75}{2007}{032109}.

\bibitem{MFYang07}\Name{Yang M. F.} \REVIEW{Phys. Rev. B}{76}{2007}{180403
(R)}; \Name{Tzeng Y. C. and Yang M. F.} \REVIEW{Phys. Rev.
A}{77}{2008}{012311}; \Name{Tzeng T. C., Hung H. H., Chen Y. C., and Yang M.
F.} \REVIEW{Phys. Rev. A}{77}{2008}{062321}.

\bibitem{HMKwok07}\Name{Kwok H. M., Ning W. Q., Gu, S. J. and Lin H. Q.} \REVIEW{Phys.
Rev. E}{78}{2008}{032103}.


\bibitem{NPaunkovic07}\Name{Paunkovi\'{c} N., Sacramento P. D., Nogueira P.,  Vieira V.
R., and Dugaev V. K.} \REVIEW{Phys. Rev. A}{77}{2008}{052302}.


\bibitem{XWang08032940}\Name{Wang X., Sun Z., and Wang Z. D.}
\REVIEW{Phys. Rev. A}{79}{2009}{012105}.



\bibitem{HMKwok08}\Name{Kwok H. M., Ho C. S., and Gu S. J.} \REVIEW{Phys. Rev. A}{78}{062302}{2008}.


\bibitem{JMa08}\Name{Ma J., Xu L., Xiong H., and Wang X.} \REVIEW{Phys. Rev.
E}{78}{2008}{051126}; \Name{Lu X. M., Sun Z., Wang X., and Zanardi P.}
\REVIEW{Phys. Rev. A}{78}{2008}{032309}.



\bibitem{AHamma07}\Name{Hamma A., Zhang W., Haas S., and Lidar D. A.} \REVIEW{Phys. Rev.
B}{77}{2008}{155111}.  

\bibitem{JHZhao0803}\Name{Zhao J. H. and Zhou H. Q.} arXiv:0803.0814.

\bibitem{SYang08}\Name{Yang S., Gu S. J., Sun C. P., and Lin H. Q.} \REVIEW{Phys. Rev. A}{78}{2008}{012304}.


\bibitem{DFAbasto08}\Name{Abasto D. F., Hamma A., and Zanardi P.} \REVIEW{Phys. Rev. A}{78}{2008}{010301(R)}.


\bibitem{HTQuan}\Name{Quan H. T. and Cucchietti F. M.} \REVIEW{Phys. Rev. E}{79}{2009}{031101}.


\bibitem{MAContinentinob}
\Name{Continentino M. A.}
  \Book{Quantum Scaling in
Many-Body Systems}
  \Publ{World Scientific Publishing, Singapore}
  \Year{2001}.





\bibitem{Lipkin1965}\Name{Lipkin H. J., Meshkov N., and Glick A. J.} \REVIEW{Nucl.
Phys.}{62}{1965}{188}; \Name{Meshkov N., Glick A. J., and Lipkin H. J.}
\REVIEW{Nucl. Phys.}{62}{1965}{199}; \Name{Meshkov N., Lipkin H. J., and Glick
A. J.} \REVIEW{Nucl. Phys.}{62}{1965}{211}.


\bibitem{JVidal05}\Name{Dusuel S. and Vidal J.} \REVIEW{Phys. Rev. Lett.}{93}{2004}{237204};
\Name{Dusuel S. and Vidal J.} \REVIEW{Phys. Rev. B}{71}{2005}{224420}.

\bibitem{SJGuQAD} \Name{Gu S. J.} arXiv:0902.4623.



\bibitem{Kitaev}\Name{Kitaev A.} \REVIEW{Ann. Phys.}{303}{2003}{2};
\Name{Kitaev A.} \REVIEW{Ann. Phys.}{321}{2006}{2}.


\bibitem{wen-book}
\Name{Wen X. G.}
  \Book{Quantum Field Theory of Many-Body Systems}
  \Publ{Oxford
University, New York}
  \Year{2004}.

\bibitem{XYFeng07}\Name{Feng X. Y., Zhang G. M., and Xiang T.}
\REVIEW{Phys. Rev. Lett.}{98}{2007}{087204}.

\end{thebibliography}
\end{document}